\documentclass[12pt,preprint]{aastex}

\usepackage{geometry}
\shorttitle{Mapping Hydrogen With Arecibo}
\shortauthors{Douglas}

\begin{document}

\pagestyle{empty}

\title{Tracing the Universe's Most Abundant Atom with the World's Largest Filled-Aperture Telescope} 


\author{Kevin A. Douglas}
\affil{University of Exeter}


\begin{abstract}

Among present-day observatories, the Arecibo Radio 
Telescope represents an extension of Galileo's vision to its logical extreme.
With a diameter of 305 metres and state-of-the-art instrumentation,
Arecibo continues to build on its legacy of world-class scientific achievement
in radio astronomy.  This paper highlights milestones in the remarkable history of
this telescope, and also discusses current surveys that are imaging the hydrogen
content of our Milky Way Galaxy, and far beyond, in unprecedented detail.

\end{abstract}




\section{Introduction}

In Galileo's most vivid dreams, could he have imagined a telescope pointed toward the heavens with a collecting
area larger than Venice's Saint Mark's Square?  In the 400 years since Galileo's work, astronomy has evolved 
enormously, and the nature of the telescopes used for observation has been an integral component of that 
revolution.  Radio astronomy is a relative newcomer to the field, having started primarily in the years 
following World War II.  Yet the radio window is extremely important, as apart from optical wavelengths it is
the only part of the electromagnetic spectrum where celestial radiation can be readily detected at the Earth's
surface, without significant attenuation by our atmosphere.  Man-made interference is increasingly a barrier
to radio observations, but that is a topic for another day.

Whether observing in the optical, radio, or any other wavelength domain, two important truths necessitate that
bigger telescopes are better telescopes.  Firstly, the sensitivity of the telescope (often called light-gathering
power) is proportional to $D^2$, where $D$ is the telescope diameter.  Thus a telescope with a larger collecting
area is inherently more sensitive to weaker emission than a smaller telescope, given the same amount of observing
time.  Secondly, the angular resolution goes as $\theta \propto \frac{\lambda}{D}$.  High resolution implies that
the resolution element $\theta$ be as small as possible, hence a larger diameter $D$ for a given wavelength $\lambda$
will lead to better resolving power.  At radio wavelengths $\lambda$ is typically $10^5$ times greater than in the
optical, underlying the need for very large radio telescopes, which push the limits of what is achievable from an
engineering standpoint.

The Arecibo Radio Telescope is currently the largest filled-aperture telescope in the world, with a diameter of
$D = 305$ metres.  Since its construction began 50 years ago, it has represented a logical extension of the ``bigger
is better" philosophy, and its scientific achievements have borne out this vision.  Below I will provide a brief
historical overview about the conception and construction of the telescope, including its important upgrades.  
Next I will review some of the scientific discoveries made with the telescope that have greatly advanced our 
knowledge of the universe.  Finally I will describe some of the current surveys underway at Arecibo, which focus
on the atomic hydrogen content of our Galaxy and in other galaxies.

\section{Historical Overview}

The history of the Arecibo Telescope is closely tied to Cornell University, and under the auspices of the National
Astronomy and Ionosphere Center (NAIC), Cornell continues to operate the telescope today.  Engineering Professor Bill
Gordon conceived the idea for the telescope, initially to conduct studies of the Earth's ionosphere through incoherent
scattering experiments.  Hence the Arecibo Ionospheric Observatory was founded, with funding from the the Advanced
Research Projects Agency\footnote{During this conference I learned that ARPA was part of the U.S.~Defense Department, 
which prompted a short discussion on the cooperation between science and the military.}.

\begin{figure}
\centering
\includegraphics[scale=0.5,angle=0]{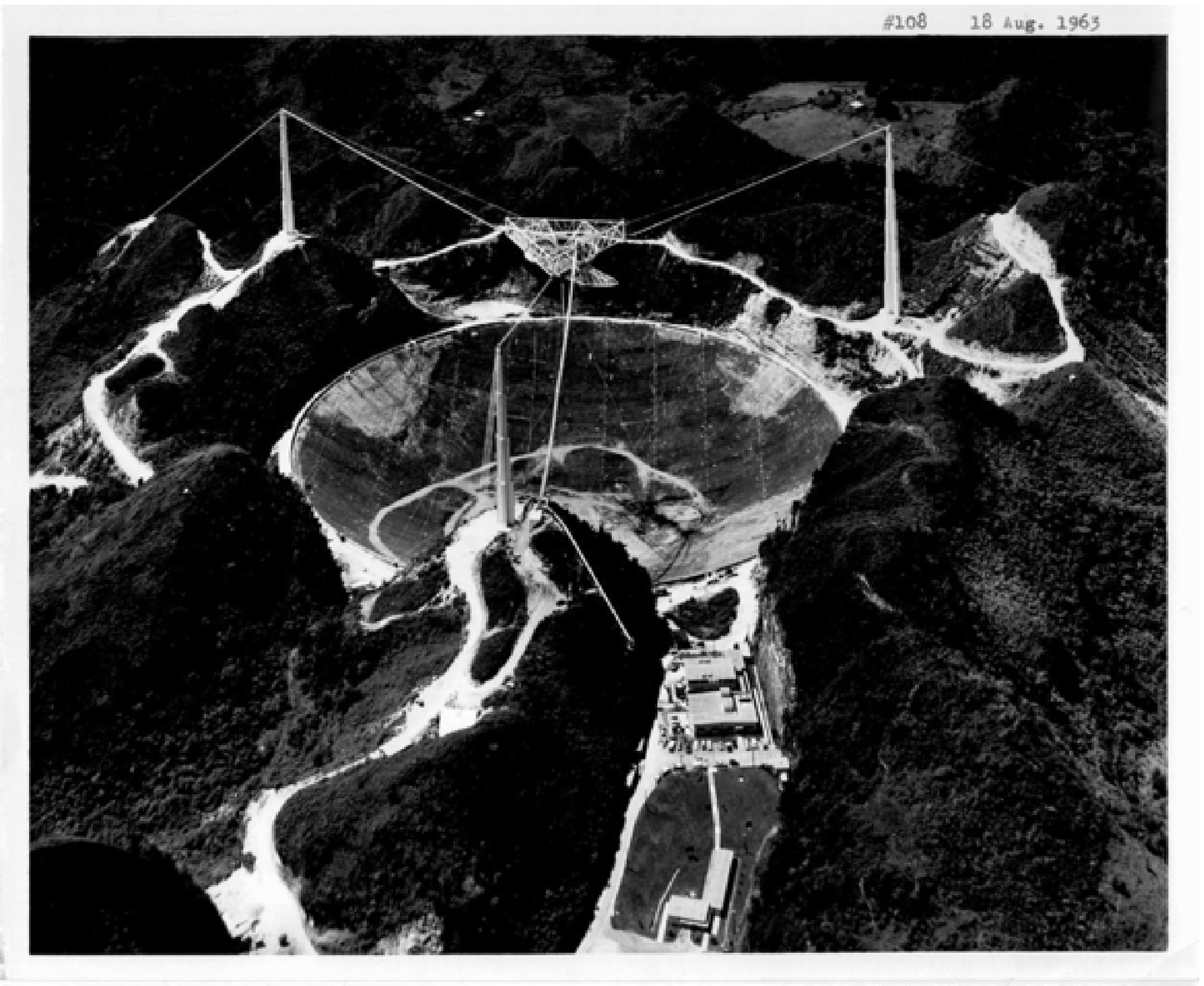}
\includegraphics[scale=0.45,angle=0]{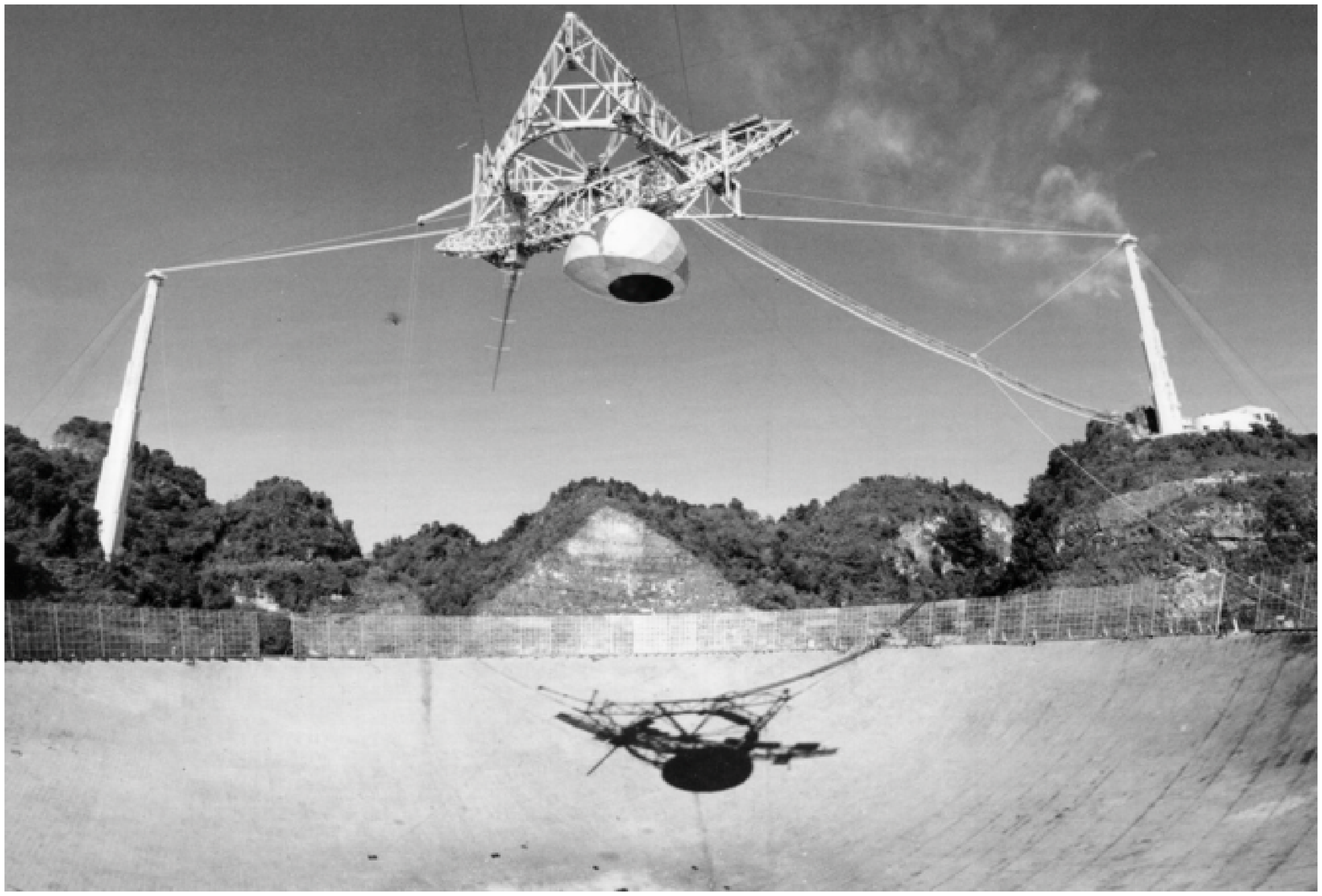}
\caption{Pictures of the Arecibo Telescope from 1963 (top), and present-day (bottom) [courtesy of NAIC]. \label{AO1963}}
\end{figure}

By 1960 the site had been chosen, and excavation had begun in the karst hills of Puerto Rico, about 15 kilometres
south of the town of Arecibo.  By 1962 the telescope's support system was in place; a roughly spherical bowl had been
carved out of the limestone topography, and three large pillars towered high above the chasm, vertices of an enormous 
equilateral triangle.  These pillars would hold up the receiver platform, an integral part of this unique telescope
design.  With no ability to steer the primary ``mirror", the receiver feeds attached to the platform could be moved
to desired coordinates by rotating a circular wheel and sliding the feed along a meridional track.  By August 1963,
the Arecibo Telescope's construction was complete, its primary reflecting surface simply a sparse mesh of wire, but
a sufficient mirror at the wavelengths required to study the ionosphere.

The potential of Arecibo for radio and radar astronomy (ie.~looking beyond our own atmosphere) pursuits was recognised
by Professor Tommy Gold, and to fulfill the promise of such research opportunities, Cornell invested in more astronomy
faculty appointments.  Moreover, this advancement in Arecibo's role required a significant upgrade to its primary
reflector.  In 1973 the telescope surface was covered with a new, denser layer of aluminum mesh, which allowed the
telescope to reflect and focus celestial radio waves from the universe toward the receivers.

In the 1990s a second major upgrade was undertaken to improve the telescope's performance.  The first phase of this
upgrade was the construction of a ground screen around the edge of the telescope, built to reduce the spillover 
noise as unwanted radio signals were reflected into the telescope from the surrounding hills.  The second phase
to the upgrade concerned the spherical shape of the primary reflector.  The ideal shape for the primary is that of
a parabola (indeed most telescope mirrors are parabolic), since all incoming rays are focussed to a point.  With
a spherical surface, the waves are instead focussed along a {\it line}, requiring the use of complicated line feed
receivers at Arecibo, which are quite lossy by nature.  Another way to compensate for this spherical aberration is
through the use of optics; a secondary and tertiary reflector were designed to focus the waves to a point.  These
reflectors were housed in a specially constructed ``Gregorian" dome, so named after the off-axis Gregorian optics
used in the upgrade.  The dome was installed in 1997, and a rotatable turret inside the dome allows many different
receivers operating at different wavelengths to benefit from the better-focussed radio emission.
Figure \ref{AO1963} shows the Arecibo Telescope in its early days, as well as a modern-day photograph.

In 2004 another upgrade was achieved with the installation of the Arecibo L-band Feed Array (ALFA), a seven-beam
dual-polarisation receiver, which essentially gives fourteen independent views of the sky at once.  Hence this
``camera" has greatly improved the mapping capability of the telescope, and the surveys mentioned below benefit
from this technological advance.

\section{Major Scientific Achievements}

The relatively short history of the Arecibo Telescope is nonetheless filled with incidents of significant 
discoveries and observations that have resulted in awards and accolades for the scientists who initiated them.
In the conference I highlighted a small number of those achievements, with which prestigious awards are 
associated.  This is not to diminish the greater number of outstanding discoveries made with Arecibo that
may not have (as of yet) been peer-recognised as award-worthy.


In 1974, rather shortly after the first telescope upgrade and in the same year that Anthony Hewish shared his 
Nobel Prize in Physics
for discovering the pulsar\footnote{Hewish shared the prize with Martin Ryle, who developed the technique of aperture
synthesis.}, Russell Hulse and Joseph Taylor of Princeton University discovered the binary motion of the pulsar
PSR 1916+13 \citep{hul75}.  A major result from this discovery was that the orbital period was slowly decaying---this finding 
came from the synthesis of many follow-up observations, {\it not} simply from the 1974 discovery.  Moreover, the
rate of decay was in precise agreement with the prediction from Einstein's General Theory of Relativity, which
suggested that such systems would lose energy as the result of the emission of gravitational radiation.  For this
work, Hulse and Taylor received the 1993 Nobel Prize in Physics.


In 1989, the Henry Draper Medal of the U.S.~National Academy of Sciences was awarded to Riccardo Giovanelli and
Martha Haynes of Cornell University, for their work in demonstrating the filamentary nature of the Pisces-Perseus
Supercluster \citep{hay88}, delineated through spectroscopic studies undertaken with the Arecibo Telescope.  Only with Arecibo's
high sensitivity and advanced signal processing capabilities was such an investigation possible.  Giovanelli and
Haynes continue to use Arecibo to trace large-scale structure in the universe.


The existence of planets orbiting other stars is an idea that has been around since Galileo's time.  Indeed the
account of Giordano Bruno's life, some of which was presented during the visit to the Venice State Archives during 
our conference, shows how controversial such views were at the turn of the 17$^{\rm th}$ century.

Many sources erroneously cite 51 Pegasi b as the first planet detected beyond the Solar System, in 1995.  It was 
indeed the first confirmed extrasolar planet orbiting a ``normal" star, but in 1992, Alex Wolszczan and Dale Frail
used precise pulsar timing measurements at Arecibo to detect the first planets beyond our Solar System \citep{wol92}.  
Based on very careful measurements, the motion of the pulsar can be explained by no less than three planets in rather 
tight orbits around the neutron star.  Two of these planets are at about the distance of Mercury to our own Sun,
while the third, smaller planet is even closer.  The American Astronomical Society (AAS) awarded Wolszczan the 
1996 Beatrice Tinsley Prize for this discovery.


While radio astronomy has probed the far reaches of the universe, Arecibo's {\it radar} astronomy programs have 
provided an incredible wealth of information about the planets and smaller bodies of our own Solar System.  These
include precise measurements of the rotation periods of Mercury and Venus, and direct imaging of the surfaces
of terrestrial planets (even Venus, through its thick, hazy, poisonous atmosphere) and asteroids.

Among the many discoveries attributed to radar astronomy with Arecibo, Jean-Luc Margot was recognised by the AAS
Division of Planetary Sciences for his studies of binary asteroids and planetary spin states.  He was awarded the
2004 Harold C.~Urey Prize for this significant contribution to planetary astronomy \citep{mar04}.

\section{Current Surveys of Atomic Hydrogen with Arecibo}

With the installation of the ALFA receiver described above, a new role for Arecibo was assumed: that of a survey
telescope.  Prior to ALFA, the telescope could be regarded as a ``single-pixel" camera, which makes large-area
surveys overly time-consuming and inefficient.  By contrast, with ALFA in operation the large majority of radio
astronomy time allocated to the telescope is consumed by mapping projects employing ALFA to survey the atomic
hydrogen content of the universe.

\begin{figure}
\centering
\includegraphics[scale=0.65,angle=0]{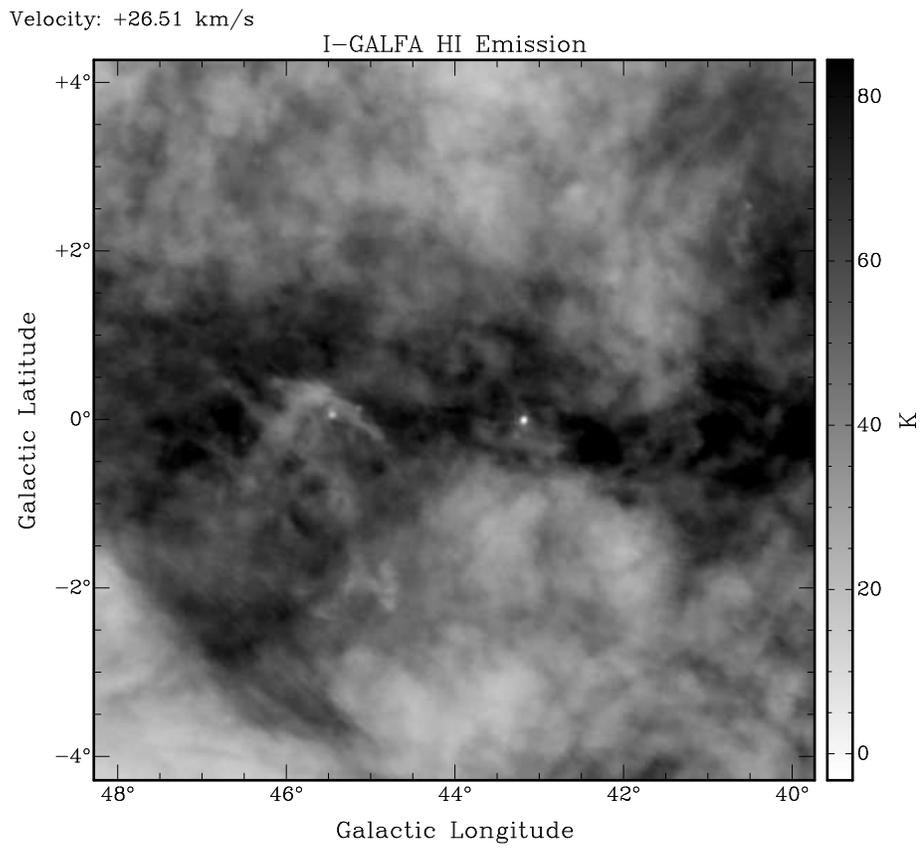}
\caption{An HI channel map from the I-GALFA survey. \label{IGAL}}
\end{figure}

Hydrogen is the most abundant atom in the universe, and in its atomic neutral form (shorthanded as HI), its use
as a tracer of interstellar structure is one of the most valuable tools in the radio astronomer's arsenal.  A
hyperfine (spin-flip) transition of HI in its ground state produces radio emission at a rest frequency of 1420.4
MHz, known widely as the 21-centimetre line of neutral hydrogen.  This frequency is centred within ALFA's ({\bf L}-band)
detection capabilities, and so the HI surveys being conducted at Arecibo with ALFA are providing new views of
the hydrogen in our Galaxy, as well as in galaxy clusters well beyond the Milky Way, with unprecedented sensitivity
and detail.

Extragalactic observations with ALFA (E-ALFA for short) seek to map the hydrogen content of any galaxies beyond
the local neighbourhood.  Giovanelli and Haynes (mentioned above) lead a project called ALFALFA \citep{gio05}, 
the Arecibo Legacy
Fast ALFA survey, which seeks to cover about half of the entire sky observable with the Arecibo Telescope (a total of
about 7000 square degrees, mainly avoiding areas where the Milky Way bands are seen).  A smaller, more targeted
survey is called AGES (Arecibo Galaxy Environment Survey), which concentrates on selected regions with a diverse
range of environments, from isolated galaxies and voids, to galaxy pairs and richer clusters.

Partnered closely to these extragalactic projects is the Galactic component of these surveys, the GALFA surveys.
GALFA-HI \citep{sta06,put09} specifically refers to the hydrogen mapping projects; there are other programs as well (eg.~continuum and
radio-recombination lines).  We take data comensally with the aforementioned E-ALFA projects, as well as conducting
our own observations in a different, much faster, observing mode.  We recently completed a survey of the Inner
Galactic Plane, called I-GALFA \citep{koo10}, which greatly supercedes previous surveys in terms of spatial and frequency coverage,
and of course sensitivity, which in turn is leading to a clearer view of the structure of our Galaxy (see Figure \ref{IGAL}).

\section{Summary}

Since its beginnings, the Arecibo Telescope has both demonstrated and achieved its potential as a unique, world class
telescope.  In the relatively young (less than 70 years) field of radio astronomy, it has already established a
legacy of outstanding scientific achievement, and it continues to build on that foundation through continued
upgrades.  Numerous innovations have been implemented, ensuring that the telescope stays current with advances in
technology.

Does this upgrading of telescopes impact their candidacy as a UNESCO World Heritage Site?  I don't believe it should.
It won't be for me to decide, but I am supremely confident that one day the Arecibo Telescope will be recognised as such.
For now, it has many years ahead as a functional astronomy facility, though we should not take it for granted.  Recent
budgetary measures within the National Science Foundation required the NAIC to reduce significantly its operational
budget for Arecibo, which resulted in the loss of many staff members, including scientists and engineers, as well as
a reduction in the number of supported receivers available to telescope users.  Nevertheless, in the opinions of many
scientists, the public, and someday UNESCO, the Arecibo Telescope represents a logical and elegant extension of
the celestial investigation begun with Galileo and others of his time 400 years ago.





\acknowledgments

The research leading to these results has received funding from the European Community's
Seventh Framework Programme under grant agreement n$^{\rm o}$ PIIF-GA-2008-221289.
I wish to thank B.M.~Lewis for his assistance with the historical background.
The Arecibo Observatory is part of the National Astronomy and Ionosphere Center, which
is operated by Cornell University under a cooperative agreement with the National
Science Foundation.




\end{document}